# *Evaluating Intersectional Fairness across Clinical Machine Learning Use Cases using Fairlogue and the All of Us Research Program*

**Nick Souligne M.S., Vignesh Subbian, PhD**
**College of Engineering, The University of Arizona, Tucson, AZ, United States**

**Abstract**

Intersectional biases in healthcare data can produce compound disparities in clinical machine learning models, yet most fairness evaluations assess demographic attributes independently. *FairLogue*, a toolkit for intersectional fairness auditing, was applied across multiple clinical prediction tasks to evaluate disparities across combined demographic groups. Using the *All of Us* dataset, two published models were selected for replication and evaluation: (A) prediction of selective serotonin reuptake inhibitor associated bleeding events and (B) two-year stroke risk in patients with atrial fibrillation. Observational fairness metrics were computed across race, gender, and intersectional subgroups, followed by counterfactual analysis to evaluate whether disparities were attributable to group membership. Intersectional evaluation revealed larger disparities than single-axis analyses; however, counterfactual diagnostics indicated that most observed disparities were comparable to those expected under randomized group membership. These results highlight the importance of intersectional fairness auditing and demonstrate how *FairLogue* provides deeper insight into bias in clinical machine learning systems.

**Introduction**

Health disparities rarely arise from a single social identity; instead, they emerge from complex interrelationships of multiple identities such as race, gender, age, and socioeconomic status, creating compounded forms of bias that can become embedded within healthcare data and the machine learning models trained on them[1]. This concept, commonly described as intersectionality, recognizes that overlapping social identities shape how individuals experience structural inequities within healthcare systems[2]. Concerns regarding these biases have become increasingly prominent as machine learning approaches are applied in clinical settings. Previous work has shown that not only do factors such as social stigma, inconsistent diagnostic labeling, and structural disparities in healthcare significantly shape the quality of healthcare data but that these societal inequities that become embedded in the data used to train these models can lead to significant model performance biases along demographic lines that may perpetuate or even reinforce these systemic inequities[3,4,5,6,7]. Furthermore, if not appropriately handled, these biases may limit the generalizability of the model to new datasets and contexts as demonstrated by prior work in a variety of clinical tasks such as survival analysis, disease progression, and even diagnostic predictions[8,9,10].

Despite growing recognition of these challenges, many fairness evaluations assess disparities along a single demographic dimension such as race or gender independently. These approaches may fail to capture the often-compounded inequities experienced by individuals that belong to multiple underrepresented groups, thereby obscuring meaningful patterns of disparity[2]. Intersectionality based approaches recognize that the experiences and outcomes for each individual are shaped by these overlapping social identities, and failure to account for these interactions may lead to misleading conclusions regarding model fairness[11]. There is a small, but growing body of research suggesting that intersectional techniques not only can improve model performance for individual subgroups but also maintain high overall predictive performance[2,11,12]. To augment these studies and build upon the understanding of these intersectional interactions, new analytical approaches that explicitly account for these relationships must be made standard practice.

Recently, a modular toolkit called *Fairlogue*[13] was made publicly available that operationalizes intersectional fairness evaluation through a combined approach of observational and counterfactual bias assessment within machine learning workflows. While initial evaluations demonstrated the feasibility of this approach within a single clinical use case, broader validation across a wider range of clinical scenarios remains a priority to understand how intersectional fairness diagnostics behave under a variety of data conditions such as outcome prevalence or demographic compositions. The goal of this present study is to evaluate the *Fairlogue* toolkit across multiple distinct clinical use cases. Specifically, this study aims to assess the generalizability, interpretability, and practicality of intersectional fairness auditing using *FairLogue* in real-world healthcare applications. Through examination of a range of predictive tasks and patient populations, this study will assess how *Fairlogue* analysis reveals context-dependent disparities and will provide empirical guidance for practitioners seeking to incorporate intersectional fairness auditing into their own clinical machine learning pipelines.

## Methods

This study utilized the *All of Us Research Program* Registered Tier V8 dataset, which is uniquely suited for algorithmic fairness investigation due to its focus on recruiting traditionally underrepresented populations and the overall scale of the dataset suggesting sizable intersectional subgroup populations. The clinical use cases are derived from prior work published using *All of Us* and include (A) a selective serotonin reuptake inhibitor (SSRI) associated bleeding event prediction task[14] and (B) a stroke risk prediction task in participants with atrial fibrillation (AF)[15]. All analysis was done in Python on the Researcher Workbench with appropriate subgroup masking procedures as defined by the *All of Us* Data and Statistics Dissemination Policy[16]. Terminology and reporting practices utilized throughout the study were informed by the AMIA Inclusive Language and Context Style Guidelines[17] to ensure respectful, context-appropriate descriptions of demographic characteristics and health disparities.

For each clinical use case, the model and cohort were replicated using electronic health record (EHR) data and survey questionnaire data. Upon validation of successful replication, the *Fairlogue* package was applied to first assess disparities along single-axis metrics and intersectional axis metrics. To quantify the overall bias associated with the intersectional demographics and to assess if these disparities were in fact driven by the group membership, the disparities were quantified under the counterfactual scenario of randomized group membership and compared to the true observed disparities. Quantification of these disparities involved the fairness metrics Equalized Odds (EO), Demographic Parity (DI), Equal Opportunity Difference (EOD), and Unfairness values ("u-values") under an acceptable threshold of 0.10 (10%) unfairness. Model performance metrics accuracy and area under the receiver operating characteristic curve (AUROC) are also reported.

*Case Study A:* The SSRI-associated bleeding events prediction task stratified patients by exposure to SSRIs including citalopram, escitalopram, fluoxetine, and sertraline. Each SSRI was modeled separately and included in a combined SSRI exposure cohort. A total of 88 features were included in the model and were comprised of sociodemographic, lifestyle, comorbidities, and medication use information. The bleeding event outcomes were defined by EHR concepts within 90 days of exposure to an SSRI and were predicted using both a light gradient boosting machine (LightGBM) and a random forest model.

*Case Study B:* The 2-year stroke risk in participants with atrial fibrillation task was particularly relevant for validation purposes due to the implementation of a disparity measure based tuning criteria based on the difference in AUROC among subgroups, and a post-processing calibration technique that set model prediction thresholds for each subgroup according to Youden's J statistic. This implementation of bias mitigation techniques further demonstrates the utility of *Fairlogue* in detecting disparities that seemingly handled by single-axis techniques. Participants were defined as those with AF and 18 years of age or older. A development or training set was populated with participants from states with lower Black/African American AF population proportions (WI, MN, AZ, MA, PA, and TX) with a test set containing patients from states with higher proportions (FL, MI, GA, IL, AL). Index dates were set as the end date of first patient visit from January 1st, 2018, to December 31st, 2019, for the development set and January 1st, 2020, to December 31st, 2021, for the test set. The outcome variable was the occurrence of a stroke within 2 years following the index date. Conditions were defined using EHR data and LightGBM was used as the prediction model. The model was tuned using participants from the training set from PA and TX accounting for approximately 30% of the training data.

## Results

### *Case Study A: SSRI-Associated Bleeding Events*

The overall cohort consisted of 12,850 participants, with 667 (5.2%) experiencing a bleeding event within 90 days of an SSRI exposure. The gender distribution was 26% Male and 74% Female, with the majority (86.8%) being White. The full demographic breakdown can be seen in Table 1.

**Table 1.** Demographic information for Case Study A – SSRI Associated Bleeding Event Predictions

| **Characteristic** | **Category** | **Citalopram n = 3,970** | **Escitalopram n = 3,511** | **Fluoxetine n = 4,178** | **Sertraline n = 5,391** | **Combined SSRIs N = 12,850** |
|---|---|---|---|---|---|---|
| Age (years) | Mean (SD) | 47.9 (14.2%) | 48.4 (15.5%) | 45.8 (14.6%) | 47.0 (15.6%) | 48.0 (15.1%) |
| Gender | Female | 2,938 (74%) | 2,576 (73.4%) | 3,184 (76.2%) | 3,935 (73%) | 9,378 (73%) |
| | Male | 1,032 (26%) | 935 (26.6%) | 994 (23.8%) | 1,456 (27%) | 3,472 (27%) |
| Race | White | 3,446 (86.8%) | 2,950 (84.0%) | 3,516 (84.2%) | 4,293 (79.6%) | 10,490 (81.6%) |
| | Black or African American | 481 (12.1%) | 505 (14.4%) | 607 (14.5%) | 1,025 (19%) | 2,177 (16.9%) |
| | Asian | 43 (1.1%) | 56 (1.6%) | 55 (1.3%) | 73 (1.4%) | 183 (1.4%) |
| Ethnicity | Hispanic | 40 (1%) | 48 (1.4%) | 63 (1.5%) | 70 (1.3%) | 167 (1.3%) |
| Bleeding Outcome | — | 199 (5%) | 191 (5.4%) | 233 (5.6%) | 272 (5.0%) | 667 (5.2%) |

Upon replication of the cohort, *Fairlogue* was applied to assess the observed disparities under single-axis metrics and intersectional metrics for both the LightGBM and random forest model. The results for the LightGBM are seen in Figure 1 and for the random forest model in Figure 2.

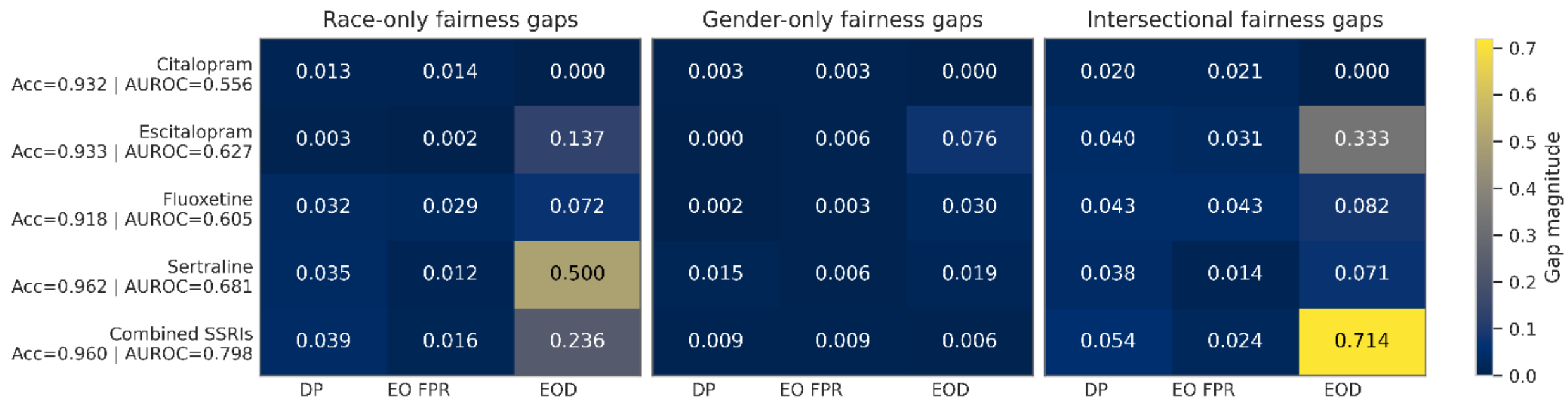

**Figure 1. Fairness comparison across demographics for SSRI-associated bleeding prediction (LightGBM models)** Heatmaps show demographic parity (DP), equalized odds false positive rate (EO FPR), and equal opportunity difference (EOD) gaps for each SSRI cohort. Results are shown separately for single axis race and gender only, and intersectional (race x gender) subgroups. Color intensity reflects the magnitude of the fairness gap, with lighter colors indicating larger disparities. Overall model performance for each cohort is displayed under the cohort labels. Consistently higher gaps are revealed in intersectional analysis, highlighting that single-axis evaluation does not capture full depth of disparities.

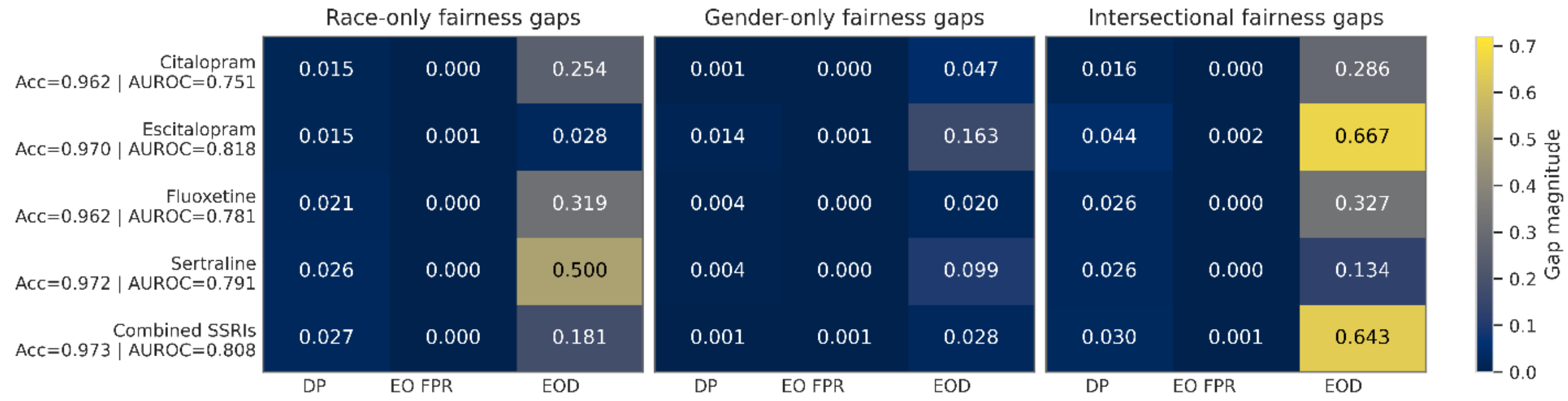

**Figure 2. Fairness comparison across demographics for SSRI-associated bleeding prediction (random forest models)** Heatmaps show demographic parity (DP), equalized odds false positive rate (EO FPR), and equal opportunity difference (EOD) gaps for each SSRI cohort. Results are shown separately for single axis race and gender only, and intersectional (race x gender) subgroups. Color intensity reflects the magnitude of the fairness gap, with lighter colors indicating larger disparities. Overall model performance for each cohort is displayed under the cohort labels. Consistently higher gaps are revealed in intersectional analysis, highlighting that single-axis evaluation does not capture full depth of disparities.

Across both modeling approaches, demographic parity and equalized odds false positive rate gaps remained small across race, gender, and intersectional subgroups, indicating minimal differences in predicted positive rates and false positive rates between demographic groups. In contrast, equal opportunity difference showed substantially larger disparities across several cohorts, reflecting variability in true positive rates between subgroups. This pattern was consistent across both LightGBM and random forest models. Although the random forest models achieved higher predictive performance overall (AUROC 0.751–0.818 compared with 0.556–0.798 for LightGBM), the relative fairness patterns remained similar, with EOD representing the primary source of disparity across both single-axis and intersectional subgroup analyses.

Taken together, these results demonstrate that although overall disparities in predicted positive rates and false positive rates remain small across both modeling approaches, meaningful differences in true positive rates persist across demographic groups. These disparities are most pronounced when evaluating intersectional subgroups, suggesting that analyses limited to single demographic axes may be underestimating potential fairness concerns within this predictive context. To assess the impact of the intersectional group membership on these disparities, *Fairlogue* counterfactual analysis was performed, and disparities are assessed under a counterfactual scenario where intersectional group membership is randomly assigned. The aggregate average, maximum, and variational counterfactual disparities are shown in Figure 3.

Counterfactual analysis revealed minimal unfairness associated with positive predictions across all cohorts. The average positive unfairness values were zero for each SSRI cohort, and no maximum or variational positive unfairness was observed. In contrast, under an acceptable threshold of 10% unfairness, modest levels of unfairness were observed for negative predictions, with average negative u-values ranging from 0.14 to 0.315 across each cohort. Similarly, maximum negative u-values showed unfairness above the threshold but with significantly elevated levels for all cohorts with the exception of sertraline. These results suggest that the models treat each intersectional subgroup similarly when predicting positives, but there is consistent disparity in negative predictions with larger maximum negative values suggesting specific subgroup outliers. These outliers can be assessed through analysis of the individual subgroup error rates in Figures 4 and 5.

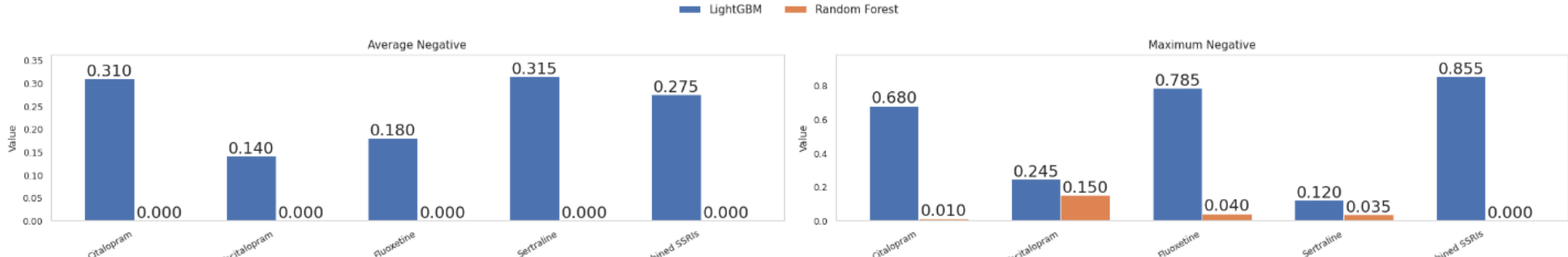

**Figure 3. Unfairness values across SSRI cohorts.** Unfairness values ("u-values") estimated using the single-robust estimator for each cohort, with an acceptable unfairness threshold of 0.1. Across all cohorts, positive-side and variance u-values were 0.0 for both models and are not shown. The only non-zero value was the variance-negative u-value for the LightGBM model in the combined SSRI cohort (0.20)

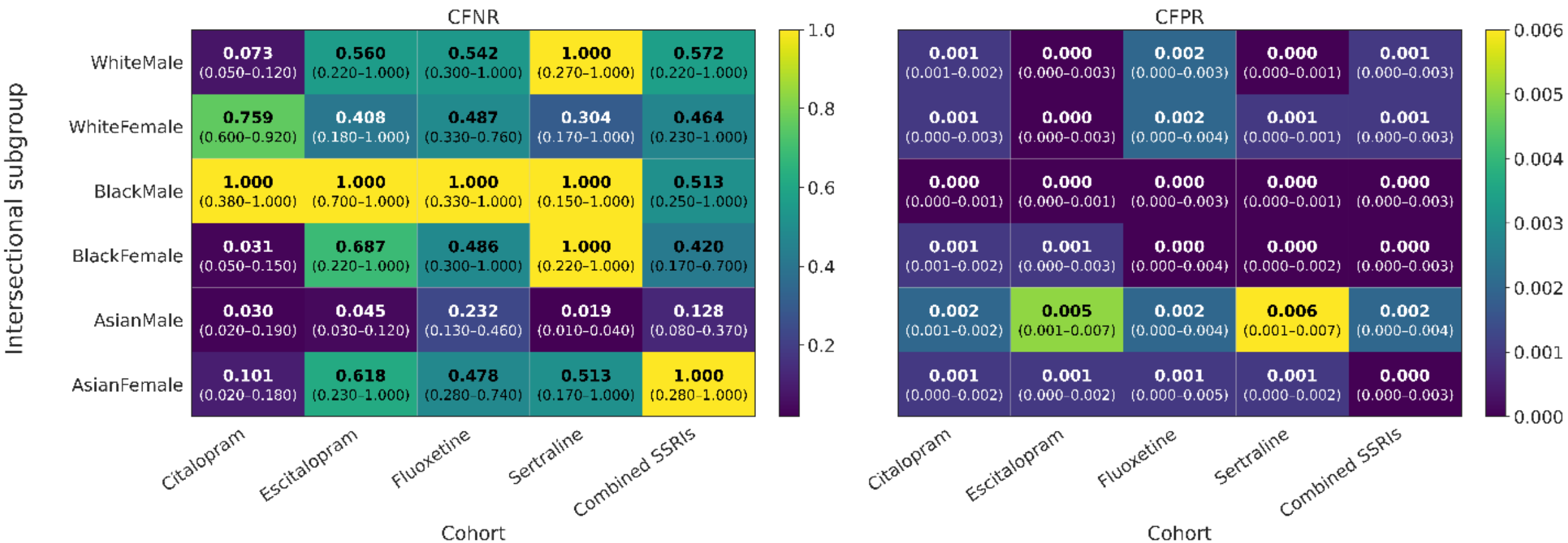

**Figure 4. Counterfactual error rates across intersectional subgroups for the LightGBM model.** Estimated counterfactual false negative rates (cFNR) and counterfactual false positive rates (cFPR) are shown with confidence intervals for each intersectional race–gender subgroup. Error rates were calculated under the counterfactual scenario of randomized group membership. Across cohorts, cFPR values remained consistently low across subgroups, while cFNR values demonstrated greater variation. However, substantial overlap in confidence intervals suggests that subgroup differences in error rates were not statistically distinct.

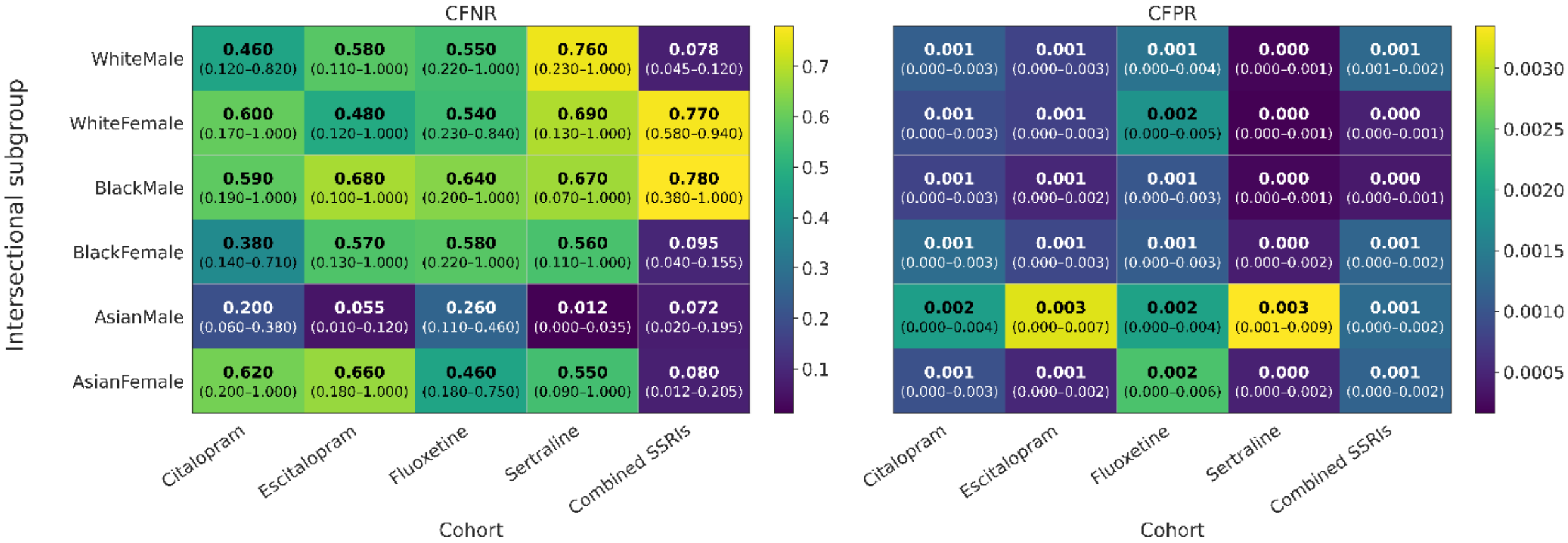

**Figure 5. Counterfactual error rates across intersectional subgroups for the random forest model.** Distribution of counterfactual false negative rates (cFNR) and false positive rates (cFPR) across intersectional demographic groups with associated confidence intervals. Compared with the LightGBM model, the Random Forest model exhibited a more uniform distribution of cFNR across subgroups while maintaining similarly low cFPR values. Overlapping confidence intervals across groups indicate that observed differences in error rates are not statistically distinct

Subgroup analysis examined counterfactual false negative (cFNR) and false positive (cFPR) rates under the counterfactual scenario of randomized intersectional group membership. Across all cohorts and both models, cFPR values were consistently minimal with little variation between subgroups while cFNR values demonstrated greater variance. The LightGBM model revealed significantly higher cFNR values across subgroups, while the random forest model showed a more uniform cFNR distribution although similar relative patterns were observed. Confidence intervals for both cFNR and cFPR overlapped substantially across demographic subgroups within each cohort. This indicates that, despite variation in error rate estimates, the observed differences are not statistically distinct, and no subgroup consistently exhibited higher counterfactual error rates across cohorts.

### *Case Study B: Stroke risk prediction in Atrial Fibrillation Participants*

Demographically, the cohort contained 11,160 participants with the majority being White (86.67%). The gender distribution was primarily Male (52.64%), and those diagnosed with a stroke within 2 years of the index date comprised 8.75% of the overall cohort. The demographic distribution of the development set followed a similar pattern with 90.78% White, and 52.89% Male with the test set containing a slightly more representative cohort consisting of 71.51% White, and 51.74% Male. The distribution of stroke outcomes was found to be 8.75% of the full cohort, with the training and testing set containing 7.11% and 14,81% respectively. These distributions can be seen in Table 2.

**Table 2.** Demographic information for Case Study B: 2-year Stroke Risk Prediction in AF participants

| Characteristic | Category | Development Set | Testing Set |
|---|---|---|---|
| States Included | — | WI, MN, AZ, MA, PA, TX | FL, MI, GA, IL, AL |
| Total Cohort Size | — | 8,777 | 2,383 |
| Age (years) | Mean (std.) | 67.45 (10.75) | 67.48 (10.97) |
| Race | White | 7,968 (90.78%) | 1,704 (71.51%) |
| | Black or African American | 809 (9.22%) | 679 (28.49%) |
| Gender | Male | 4,642 (52.89%) | 1,233 (51.74%) |
| | Female | 4,135 (47.11%) | 1,150 (48.26%) |
| Stroke Outcome | — | 624 (7.11%) | 353 (14.81%) |

After replicating the cohort, *Fairlogue* was applied to assess observational disparities along single-axis and intersectional race and gender demographics for the full cohort. Overall observed disparities were minimal for both single-axis and intersectional demographics after applying the tuning criteria based on subgroup AUROC differences. Race-only disparities for DP, EO FPR, and EOD were 0.131, 0.101, and 0.018 respectively. Gender-only disparities were similarly minimal across the board with DP at 0.015, EO FPR at 0.013, and EOD at 0.032. When evaluating the intersectional subgroups, disparities were slightly elevated in comparison to gender only metrics but still relatively low with DP at 0.138, EO FPR at 0.103, and EOD at 0.094. Overall model performance remained moderate, with an accuracy of 0.817 and AUROC of 0.759. These results are compiled below in Table 3.

**Table 3.** Observed disparities for single-axis demographics (race and gender) and intersectional demographics (race/gender)

| Metric | Fairness Metric (Race only) | Fairness Metric (Gender only) | Intersectional Metric |
|---|---|---|---|
| Demographic parity gap | 0.131 | 0.015 | 0.138 |
| Equalized odds FPR gap | 0.101 | 0.013 | 0.103 |
| Equal opportunity gap | 0.018 | 0.032 | 0.094 |
| *Model Accuracy: 0.817; **Model AUROC: 0.759 | | | |

Similar to the first use case, *Fairlogue* counterfactual analysis was performed, and disparities were assessed under a counterfactual scenario where intersectional group membership is randomly assigned. Using the counterfactual framework with an acceptable threshold of unfairness set to .1, the aggregated u-value metrics all approached zero indicating minimal unfairness attributable to intersectional group membership. The distribution of the discrepancies between observed disparities and the counterfactual scenario are seen in Figure 6.

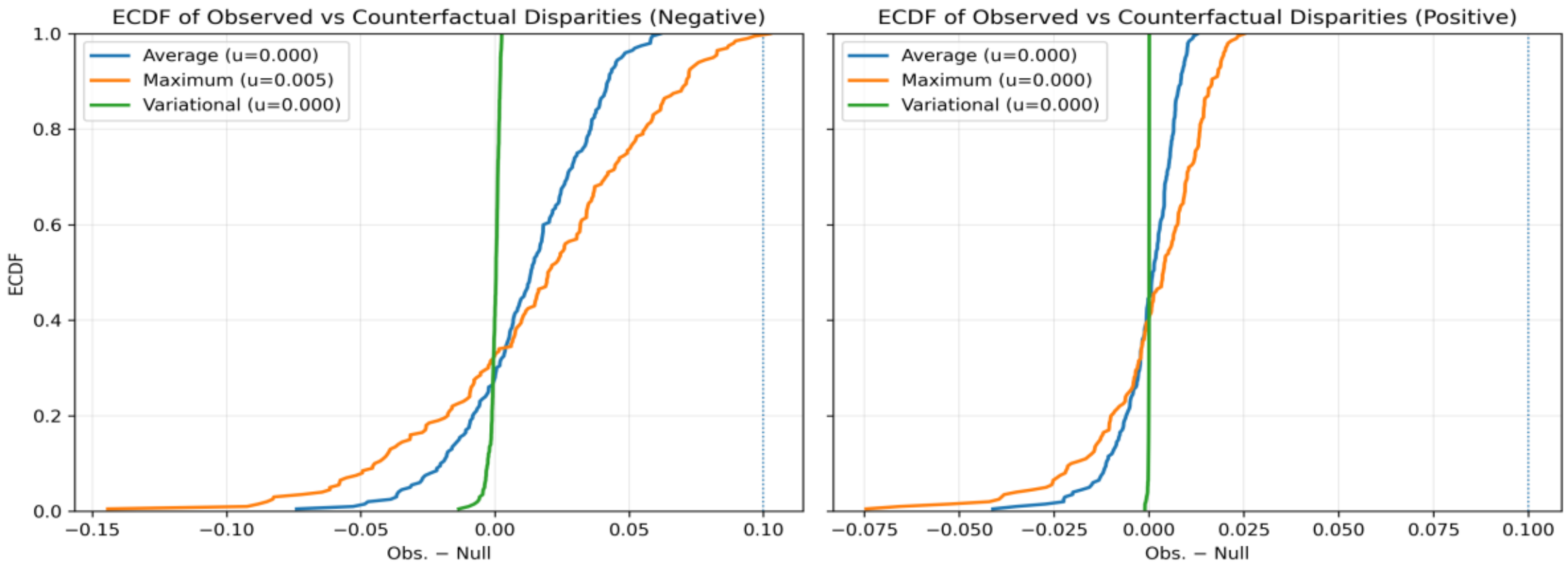

**Figure 6**. **Empirical cumulative distribution of disparity differences under counterfactual analysis for the stroke prediction task** Empirical cumulative distribution functions (ECDFs) show the difference between observed fairness disparities and their corresponding counterfactual disparities across aggregate unfairness metrics. The dotted line represents the acceptable unfairness threshold (0.1). Distributions are concentrated near zero and remain well below the threshold, indicating that the magnitude of observed disparities is comparable to that expected under randomized intersectional group membership.

Across all aggregated metrics, the empirical cumulative distribution functions are concentrated near or below zero and remain well below the unfairness threshold. These results indicate that the observed disparities are comparable to those expected under the randomized group membership counterfactual scenario, and that the magnitude of disparity attributable specifically to intersectional group membership is minimal in this prediction task. Further analysis of the individual subgroup error rates shows noticeable cFNR variation between subgroups, with several groups exhibiting both higher average estimates and wider confidence intervals in the counterfactual scenarios. Contrasting these results, the cFPR values were substantially smaller overall and demonstrated less variation between groups. Across both error rates, the estimated subgroup means generally fell within the confidence intervals of each other group, indicating that the observed error rates were consistent with the values expected under randomized demographic assignment. These error rates can be seen in Figure 7.

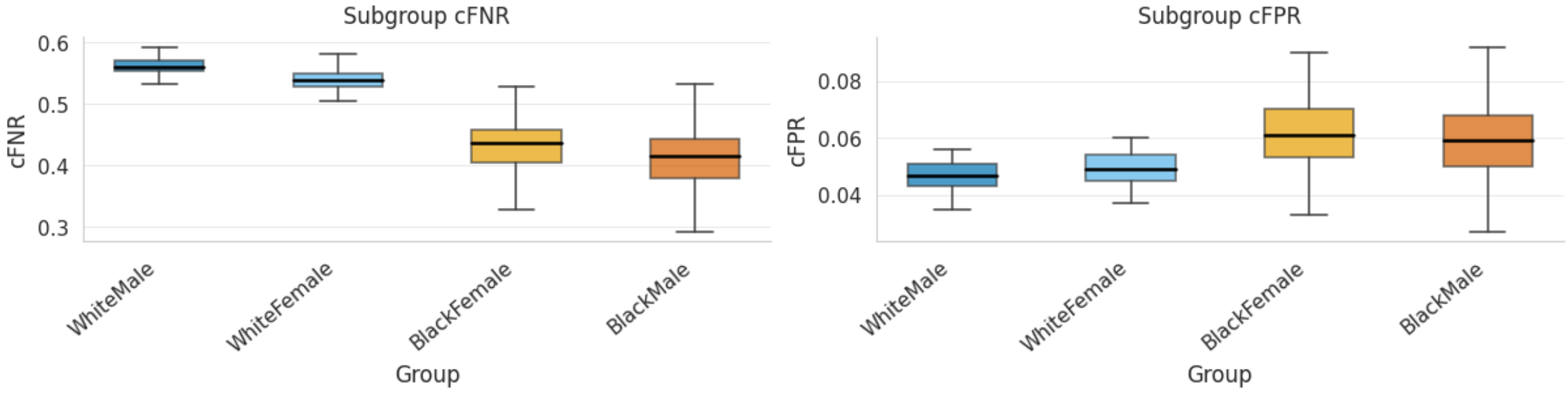

**Figure 7**. **Counterfactual subgroup error rates for the atrial fibrillation stroke prediction model.** Estimated counterfactual false negative rates (cFNR) and false positive rates (cFPR) with confidence intervals for each intersectional demographic subgroup. While variability in cFNR estimates is observed across groups, cFPR values remain consistently low. Substantial overlap in confidence intervals indicates that subgroup differences in error rates are consistent with those expected under the counterfactual scenario of randomized demographic assignment.

**Discussion**

The results of this study demonstrate how intersectional fairness auditing can provide a more comprehensive understanding of model performance across demographic groups. Both clinical use case applications of *Fairlogue* in this study revealed patterns that would not have been discovered using traditional single-axis fairness evaluations. In particular, intersectional analyses consistently exposed disparities that were either minimized or obscured when either race or gender were evaluated independently. These findings reinforce prior work suggesting that fairness assessments based solely on individual demographic attributed may be underestimating both the magnitude and complexity of model inequities across patient populations[10]. Importantly, these patterns were consistent across both LightGBM and random forest models. Although the random forest models achieved slightly higher predictive performance in several cohorts, the overall fairness patterns remained largely unchanged. Demographic parity and false positive rate disparities remained minimal, while equal opportunity difference consistently represented the largest source of disparity across subgroups. The consistency of these results across two distinct model architectures suggests that the observed fairness patterns are likely driven more by the underlying data distribution than by properties of any specific modeling approach.

In the SSRI-associated bleeding prediction task, disparities in predicted positive rates and false positive rates were minimal across demographic groups. However, equal opportunity difference demonstrates greater variability across demographic groups, indicating model disparities in predictive true positive events between each subgroup. Additionally, intersectional disparities were found to be higher than single axis demographic disparities across all metrics which further reinforces that single-axis evaluation does not reveal the full scope of model bias and performance inequities. Counterfactual analysis provided further contextualization of these observed disparities by demonstrating that the majority of observed differences were comparable to those expected under randomized group membership. These findings, taken together, suggest that while disparities are present in the observed data, they may not arise solely from demographic group identity but rather through more complex correlations between these identities and other covariates in the model, such as comorbidities, medication exposures, or other socioeconomic factors embedded in the dataset.

The second clinical use case involving 2-year stroke risk prediction in atrial fibrillation participants demonstrated similar patterns. The observational metrics indicated relatively small disparities along single-axis race and gender demographics, yet again intersectional evaluation revealed significantly larger disparities in both predictive positive rates and false positive rates along combined race and gender demographics compared to the single axis gender only disparities. However, counterfactual analysis again demonstrated that the magnitude of unfairness attributable specifically to group membership was minimal. Together, these findings suggest that intersectional disparities in these prediction tasks are likely driven by broader covariate relationships embedded in the dataset rather than direct model discrimination based on demographic characteristics.

Across both clinical use cases, extremely low false positive rates were consistently observed across demographic subgroups. This pattern is likely influenced by the relatively low prevalence of positive outcomes in the datasets, particularly in the SSRI bleeding prediction task. In such scenarios, predictive models tend to adopt conservative decision boundaries that minimize false positive predictions, which may limit the interpretability of fairness metrics based on false positive rates. As a result, disparities in true positive rates may provide a more informative signal of subgroup performance differences in low-prevalence clinical prediction tasks.

Taking both clinical applications together, these results highlight the importance of taking intersectionality into account when quantifying model disparities and distinguishing between these observed disparities arising from group membership or other factors. Observed disparities may arise from a variety of factors such as data imbalances, differences in outcome prevalence across populations, or even structural inequities reflected in healthcare. However, counterfactual analysis may provide an additional layer of diagnostic insight by assessing whether those disparities arise solely from those factors or through other spurious model features. Determining if these disparities persist under counterfactual scenarios where group membership is randomized allows for more precise bias detection, and ultimately mitigation. In the scenarios examined in this study, the counterfactual analysis suggests that the model themselves may not be explicitly encoding discriminatory prediction behavior, despite observable disparities remaining present across intersectional demographic groups. This suggests that more effective efforts of bias mitigation may require looking at different correlations between intersectional subgroups and factors such as comorbidity prevalence, or socioeconomic characteristics.

From a practical perspective, these findings reinforce and further illustrate prior work demonstrating the utility of Fairlogue in helping practitioners quantify and interpret fairness within the broader context of model development and deployment. Observational metrics alone may suggest the presence of bias, but counterfactual diagnostics can help determine whether these disparities are directly attributable to demographic group membership or arise from more complex covariate relationships. This distinction is critical for future development work of appropriate mitigation strategies, as intervention strength and downstream healthcare outcomes will be determined by either model structure or underlying data distributions.

Several limitations should be considered when interpreting these results. First, both clinical use cases were derived from the *All of Us Research Program* dataset, which, while designed to improve representation of historically underrepresented populations, still contains demographic imbalances that may influence fairness estimates. Second, the counterfactual framework relies on the assumption that the observed covariates sufficiently capture variance associated between group membership and outcome risk. Additional confounding influence by other covariates or even misspecification of the model may influence diagnostic interpretability. Additionally, the relatively low representation of positive outcomes in the data distribution itself may limit the interpretability of the fairness metrics as previously mentioned. Finally, despite application on two distinct clinical tasks were examined, further validation across a wider range of clinical applications and models would further strengthen the generalizability of these findings.

Future work should explore how *Fairlogue* can be directly integrated into model development workflows, including deployment and assessment of ongoing models, to support fairness-aware model evaluation. Additionally, further investigation into the relationships between covariates and demographic characteristics may be a particularly insightful approach that could help identify the underlying drivers of the observed disparities and help inform more targeted and effective bias mitigation strategies.

## Conclusion

This study demonstrates the practical application of the *Fairlogue* toolkit for intersectional fairness auditing across multiple clinically relevant machine learning use cases. With analysis of both observational fairness metrics and counterfactual fairness, *Fairlogue* provides a structured toolkit for identifying disparities across demographic groups and assessing whether those disparities are directly attributable to group membership. In this study, both clinical use cases demonstrated additional disparities along intersectional axes compared to single axis demographics. However, counterfactual analysis indicated that much of the observed disparities may not be directly related to the intersectional group membership itself, highlighting the importance of distinguishing the source of model bias. Together, these findings demonstrate that intersectional fairness auditing can provide deeper insight into the behavior of clinical machine learning models and support more informed decisions regarding model development, evaluation, and deployment. As machine learning continues to be integrated into clinical decision-making support systems, tools such as *Fairlogue* will play a pivotal role in ensuring that predictive models are evaluated in ways that reflect the complex and intersecting identities present within real-world patient populations.

## Acknowledgements

We gratefully acknowledge *All of Us* participants for their contributions, without whom this research would not have been possible. We also thank the National Institute of Health's *All of Us Research Program* for making available the participant data examined in this study. The *All of Us* Research Program is supported by the National Institutes of Health (NIH), Office of the Director (including award # OT2OD037642 and # OT2OD036485).